\documentclass[runningheads]{llncs}
\usepackage{amsmath}
\usepackage{booktabs}
\usepackage{multirow}
\usepackage[T1]{fontenc}
\usepackage{graphicx,verbatim}
\begin{document}
\title{LinGuinE: Longitudinal Guidance Estimation for Volumetric Tumour Segmentation}
\titlerunning{LinGuinE: Longitudinal Guidance Estimation}

%% Removed for anonymized MICCAI submission
\author{Nadine Garibli \and
Mayank Patwari \and
Bence Csiba \and Kostantinos Sidiropoulos \and Yi Wei}

\authorrunning{N. Garibli et al.}
\institute{AstraZeneca, 1 Francis Crick Avenue, Cambridge, United Kingdom \\
\email{nadine.garibli@astrazeneca.com}}

% \author{Anonymized Authors}  %% Added for anonymized MICCAI submission
% \authorrunning{Anonymized Author et al.}
% \institute{Anonymized Affiliations \\
%     \email{email@anonymized.com}}
  
\maketitle              % typeset the header of the contribution
\begin{abstract}
Longitudinal volumetric tumour segmentation is critical for radiotherapy planning and response assessment, yet this problem is under-explored and most methods produce single-timepoint semantic masks, lack lesion correspondence, and offer limited radiologist control. We introduce LinGuinE (\textbf{L}ongitud\textbf{in}al \textbf{Gui}da\textbf{n}ce \textbf{E}stimation), a PyTorch framework that combines image registration and guided segmentation to deliver lesion-level tracking and volumetric masks across all scans in a longitudinal study from a single radiologist prompt. LinGuinE is temporally direction agnostic, requires no training on longitudinal data, and allows any registration and semi-automatic segmentation algorithm to be repurposed for the task. We evaluate various combinations of registration and segmentation algorithms within the framework. LinGuinE achieves state-of-the-art segmentation and tracking performance across four datasets with a total of 456 longitudinal studies. Tumour segmentation performance shows minimal degradation with increasing temporal separation. We conduct ablation studies to determine the impact of auto-regression, pathology specific finetuning, and the use of real radiologist prompts. We release our code and substantial public benchmarking for longitudinal segmentation, facilitating future research.

% Authors must provide keywords and are not allowed to remove this Keyword section.

\end{abstract}
\section{Introduction}
Volumetric segmentation of gross tumour volumes is an important first step in radiotherapy and interventional planning \cite{Savjani2022AutomatedRadiotherapy}. Volumetric assessments also show promise for evaluating response to oncological therapy \cite{Sargent2009ValidationEnd-points}. Response is normally studied over multiple scans, acquired at different timepoints, of the same patient. 

Presently, deep learning methods achieve strong automatic segmentation of tumour volumes \cite{Primakov2022AutomatedImages}, however most approaches, including those that incorporate multiple timepoints \cite{rokuss2024longitudinal,omnimamba,Kruger2020FullyNetworks}, are designed to produce semantic segmentations of a single volume at a single timepoint. Such fully automatic methods do not perform lesion tracking, which is essential for measuring treatment response and assessing disease progression. Moreover, fully automatic methods do not allow radiologist input, such as choosing lesions of focus, which reduces the chance of clinical adoption \cite{Bernstein2023CanRadiography}. Promptable semi-automatic segmentation, popularised by Segment Anything \cite{Kirillov2023SegmentAnything} and adapted to medical imaging to incorporate radiologist guidance (e.g., points or bounding boxes) \cite{Luo2021MIDeepSeg:Learning}, typically improves performance over fully automatic pipelines and supports segmentation of individual tumour volumes. However, semi-automatic methods also operate at single timepoints, providing limited support for the analysis of an entire longitudinal study.

There exist several methods that track tumours in longitudinal studies by predicting their locations but do not produce segmentations \cite{Cai2021DeepStudies,Vizitiu2023}. To the best of our knowledge, only three methods handle simultaneous tracking and segmentation. Hering et al. \cite{hering21a} use image registration to define a region of interest (ROI) around each tumour based on a radiologist-provided location in a previous timepoint and invokes an nnUNet to generate segmentations on follow up scans \cite{Isensee2021NnU-Net:Segmentation}. Therefore, radiologist guidance is limited to an ROI in the follow ups, capping segmentation performance at that of the automatic nnUNet. LesionLocator \cite{LesionLocator} addresses this by introducing an end-to-end trainable model, combining registration and promptable segmentation for tumours on CT. However, if performance on a dataset is limited, the model cannot be finetuned without a relevant labelled longitudinal dataset with tracked lesions, which are difficult to obtain. Hein et al. \cite{Hein2025} perform automatic segmentation and assign tumour correspondence with a CNN. Hence, this method cannot leverage radiologist inputs and requires training the CNN on longitudinal data. Additionally, existing methods are evaluated in the context of a single modality and cancer type, and on private datasets with a limited number of timepoints per study \cite{Hein2025,hering21a,LesionLocator}, which prohibits in-depth analysis of temporal performance degradation.

To address these limitations and provide a well-performing out-of-the-box solution, we present LinGuinE (\textbf{L}ongitud\textbf{in}al \textbf{Gui}da\textbf{n}ce \textbf{E}stimation). LinGuinE is a customizable framework for the tracking and segmentation of tumours in longitudinal studies from a prompt provided at a single timepoint. LinGuinE combines prompt propagation using image registration with guided segmentation, and is the first framework that allows arbitrary registration and semi-automatic segmentation algorithm (with no requirement for training on longitudinal data) to be directly repurposed. We demonstrate that LinGuinE:

\begin{enumerate}
    \item Achieves state-of-the-art results from minimal radiologist input.
    \item Is the first temporally direction agnostic framework, allowing radiologists to provide prompts at any timepoint, facilitating retrospective analysis.
    \item Outperforms existing methods by combining public algorithms, making LinGuinE a state-of-the-art solution in the absence of longitudinal training data.
    \item Is more robust to temporal degradation in performance than other methods.
\end{enumerate}

To facilitate further research, our contribution includes a PyTorch package (https://github.com/n-e-garibli/LinGuinE) and the first benchmarking of longitudinal algorithms on public data, multiple cancer types, and studies with more than three timepoints. The segmentation masks generated with various configurations of LinGuinE will be made publicly available.

% We also release tumour associations for the UniToChest dataset [MAYBE NOT ACTUALLY!] \cite{unitochest} where multiple timepoints are available per patient, turning it into one of the only publicly available datasets with tumour segmentations and tracking (LINK).

\section{Methods}
\subsection{The LinGuinE Framework}
Let $S$ denote a set of volumetric scans from the same patient. Let $I_i \in S$ be a scan at timepoint $i$ and $P_{i}^{k}$ be the 3D coordinate center of a tumour $k$ on $I_{i}$. LinGuinE produces a segmentation for the same tumour on every scan in $S$,

\begin{equation}
    \textit{LinGuinE}(I_i, P_{i}^{k}) \rightarrow \{M_j(k) \;|\; I_j \in S\}
\label{linguine}
\end{equation}
 where we call $I_i$ the source scan, $I_j$ the destination scan at timepoint $j$, and $M_j(k)$ the segmentation mask of tumour $k$ in the destination scan. Any scan in $S$ can be selected as the source. Consider also a segmentation algorithm that consumes a guidance point to produce a mask at any timepoint, such that,

\begin{equation}
    \textit{GuidedSeg}(I, P^{k}) \rightarrow \ M(k) \,
\label{guided_seg}
\end{equation}

% When $i = j$, the longitudinal segmentation problem reduces to segmenting a single scan. Invoking the SwinUNETR produces the segmentation mask $M_i(k)$ on the scan $I_i$: SwinUNETR$(I_i, C_i) \rightarrow M_i(k)$. When $i \ne j$,
LinGuinE starts from $M_i(k)$, which can be provided by a radiologist directly, or obtained by invoking the segmentation algorithm from an input radiologist click, as in equation \eqref{guided_seg}. $P_{i}^{k}$ is then sampled as the center of $M_i(k)$ and is propagated from the coordinate space of $I_i$ to the coordinate space of $I_j$ using image registration. We denote a propagated point as $P_{i \rightarrow j}^{k}$. The intuition behind point propagation is that tumours maintain an anatomically similar location across timepoints. The registration algorithm is a configurable component of the framework. In this work, we evaluate LinGuinE using publicly available GradICON models (UniGradICON (UGI) \cite{uniGradICON} and, to show the customization benefit in the context of lung cancer, the LungGradICON (LGI) which is designed specifically for lung registration \cite{GradICON}). Upon computing $P_{i \rightarrow j}^{k}$, LinGuinE invokes a guided segmentation model to produce $M_{j}(k)$ on $I_j$.
% Additionally, we use Arun’s method \cite{arun1987least} for rigid registration based on anatomical landmarks derived from the centers and edges of TotalSegmentator masks \cite{Wasserthal2023TotalSegmentator:Codes}.
 \begin{equation}
    \textit{GuidedSeg}(I_j, P_{i \rightarrow j}^{k}) \rightarrow M_j(k)\\
\label{guided_seg_destination}
\end{equation}

After inference, LinGuinE keeps the connected component closest to $P_{i \rightarrow j}^{k}$ as the final $M_j(k)$ to avoid false positives in nearby tumours. The segmentation of a longitudinal study is then defined as;

\begin{equation}
    \textit{LinGuinE}\left(I_i, P_{i}^{k}\right) \rightarrow \{\textit{GuidedSeg}\left(I_j, P_{i \rightarrow j}^{k}\right) \;|\; I_j \in S\}
\label{full_linguine_eq}
\end{equation}

However, the location, shape, and size of a tumour can change over time and $P_{i \rightarrow j}^{k}$ may land outside the tumour in $I_{j}$, even with accurate registration. Hence, performance may be enhanced by leveraging the segmentation model to pull a propagated point closer to the center of a tumour with a technique called `boosting'. In boosted configurations, the final prompt to the model is the center of an initial lesion segmentation obtained using $P_{i \rightarrow j}^{k}$. The below equation then replaces equation \eqref{guided_seg_destination} to obtain $M_j(k)$. This is an optional feature in LinGuinE.

 \begin{equation}
    \textit{GuidedSeg}\left(I_j, \textit{Center}\left(\textit{GuidedSeg}\left(I_j, P_{i \rightarrow j}^{k}\right)\right)\right) \rightarrow M_j(k) \\
\label{boosted_guided_seg_destination}
\end{equation}

\subsection{Guided Volumetric Segmentation}

Any guided segmentation algorithm can be used in the LinGuinE framework. Although ROI-based guidance, as leveraged by Hering et al. \cite{hering21a}, offers coarse spatial information, point prompts encode precise locations. Recent advances in the field led to the availability of many point-prompted models for tumour segmentation, and we argue that directly repurposing them yields strong performance in a longitudinal setting. In this work, we evaluate nnInteractive \cite{isensee2025nninteractive}, Vista3D \cite{vista3D}, and the tumour segmentation component of LesionLocator (LLSeg) \cite{LesionLocator} within LinGuinE. To demonstrate the benefit of pathology specific optimization, we fine-tune the aforementioned models for lung tumour segmentation.

The datasets used for finetuning include 1) 418 CT scans from the publicly available NSCLC-Radiomics dataset \cite{Aerts2014DecodingApproach}, collected at the MAASTRO clinic in the Netherlands, 2) 63 CT volumes from task three of the Medical Imaging Decathlon \cite{msd_dataset}, and 3) A selection of 635 CT scans from a confidential phase 3 study of stage IV non-small cell lung cancer (NSCLC). From each dataset, 16\% of the scans were set aside for validation and an additional 16\% for testing, leaving 68\% for training. For each training run, the same image preprocessing as expected by the pretrained weights was used. The models were trained for up to 200 epochs on an NVIDIA A10G GPU using the PyTorch lightning framework. The checkpoint with the lowest validation Dice-CrossEntropy loss was selected.

\subsection{Test Datasets}

We evaluate LinGuinE on four datasets, unseen by any of the tested segmentation algorithms. The first is 4DCBCT \cite{Balik2013EvaluationTherapy}, consisting of CBCT studies and primary tumour segmentations from 10 NSCLC patients over the course of 9 weeks of chemoradiotherapy. Each patient has up to 9 timepoints and 10 scans per time point (one per breathing cycle phase). We use the scan at the 90\% phase, corresponding to an inhale. This defines a subset of 50 scans.

The second dataset, Phase-3, is a private dataset originating from a confidential phase 3 monoclonal antibody study targeting metastatic NSCLC. This dataset encompasses data from 30 countries and 2,398 patients. We randomly selected and annotated up to 10 lesions per patient in 354 CT scans from 95 patients, spanning over 52 weeks. Each patient has at least four timepoints segmented by two radiologists with over 15 years of combined clinical experience.

The third is autoPET Longitudinal CT (APLCT) \cite{LongitudinalCT2025}, containing whole-body CT at two timepoints across 300 studies of melanoma patients undergoing treatment. The images were acquired at the University Hospital Tübingen.

Finally, UniToChest \cite{unitochest} is comprised of 715 thoracic CT volumes from 623 lung cancer patients, collected and annotated at Citta della Salute e della Scienze Hospital. We selected 51 patients with multiple timepoints and at least one tumour larger than 5mm in axial diameter. At least one tumour per patient was then associated across time by a senior radiologist to create a test set of 78 unique lesions across 109 scans. Our associated ground truth masks will be made publicly available to support benchmarking and further research.

\section{Experimental Results}

\subsection{Segmentation Performance}

\begin{figure}[t]
    \centering
    \includegraphics[width=\columnwidth]{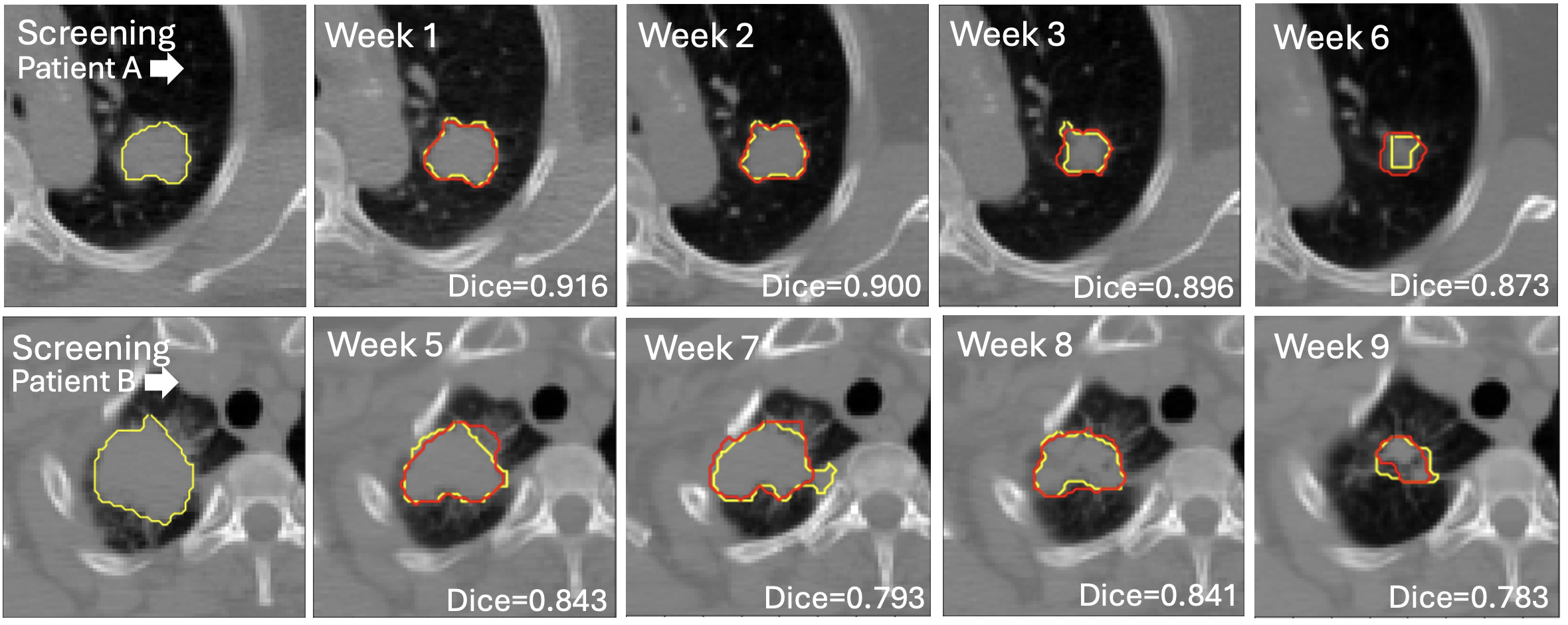}
    \caption{LinGuinE segmentations of two longitudinal studies in the 4DCBCT dataset. The center of the screening label was propagated to obtain segmentations (red contours) on subsequent timepoints. The ground truth is denoted by yellow contours.}
    \label{fig:good_examples}
\end{figure}

To demonstrate state-of-the-art performance, we propagated radiologist guidance from the baseline scan to all subsequent scans per patient as in Figure \ref{fig:good_examples}. We evaluated multiple registration–segmentation combinations; the best-performing configurations are reported in Table \ref{results_table_median}. We used Dice to evaluate overall segmentation quality and the Euclidean Distance, ED (mm), between the centers of the ground truth and predicted mask to evaluate lesion tracking. We computed metrics on a per-lesion basis, hence, mis-tracked tumours receive a Dice of zero. Therefore, the metric distributions are not Gaussian and we report the median to more fairly capture performance.

\begin{table}[t]
\centering
\caption{Median Dice and ED (mm) on the test datasets with various configurations of LinGuinE denoted as (registration, segmentation) compared to other methods. We only include a boosted configuration if it yielded the top metric for a dataset. `Finetuned' refers to models finetuned to segment lung tumours. APLCT is not evaluated with lung-specific algorithms, since it contains tumours in all parts of the body. Best metrics are in \textbf{bold}. \textit{Italics} indicates that the metric was undefined in >20\% of cases due to false negative predictions.}
\label{results_table_median}
\begin{tabular}{l@{\hspace{4pt}}cc@{\hspace{4pt}}|@{\hspace{4pt}}cc@{\hspace{4pt}}|@{\hspace{4pt}}cc@{\hspace{4pt}}|@{\hspace{4pt}}cc}
\toprule
& \multicolumn{2}{c@{\hspace{4pt}}|@{\hspace{4pt}}}{\textbf{Phase-3}} & \multicolumn{2}{c@{\hspace{4pt}}|@{\hspace{4pt}}}{\textbf{4DCBCT}} & \multicolumn{2}{c@{\hspace{4pt}}|@{\hspace{4pt}}}
{\textbf{UniToChest}} & \multicolumn{2}{c}{\textbf{APLCT}} \\
\cmidrule{2-3} \cmidrule{4-5} \cmidrule{6-7} \cmidrule{8-9}
& \textbf{Dice} & \textbf{ED} & \textbf{Dice} & \textbf{ED} & \textbf{Dice} & \textbf{ED} & \textbf{Dice} & \textbf{ED} \\
\midrule
\multicolumn{9}{l}{\textbf{Existing Method}} \\
LesionLocator & 0.164 & \textit{2.786} & 0.098 & 16.331  & 0.520 & \textit{1.770} & 0.303 & \textit{\textbf{1.035}} \\
Hering et al. & 0.570 & 9.015 & 0.728 & 5.722 & 0.606 & 5.226 & \multicolumn{2}{c}{-}  \\
\midrule
\multicolumn{9}{l}{\textbf{LinGuinE Configuration}} \\
UGI, Vista3D & 0.528 & 3.611 & 0.334 & 6.427 & 0.414 & 4.391 & 0.080 & 7.462 \\
UGI, nnInteractive & 0.596 & 3.819 & 0.719 & 4.712 & 0.556 & 5.390 & 0.023 & 14.942 \\
UGI, LLSeg & 0.745 & 1.730 & 0.680 & 4.094 & 0.642 & 5.131 & 0.433 & 4.131  \\
LGI, Vista3D & 0.539 & 3.533 & 0.350 & 4.056 & 0.629 & 2.624 & \multicolumn{2}{c}{-} \\
LGI, nnInteractive & 0.612 & 3.784 & 0.704 & 4.470 & 0.699 & 2.461 & \multicolumn{2}{c}{-} \\
LGI, LLSeg & 0.759 & 1.809 & 0.717 & 4.105 & \textbf{0.786} & 1.262 & \multicolumn{2}{c}{-} \\
LGI, Finetuned Vista3D & 0.732 & 1.917 & \textbf{0.773} & \textbf{3.125} & 0.698 & 1.876 & \multicolumn{2}{c}{-} \\
LGI, Finetuned nnInter. & 0.712 & 2.051 & 0.702 & 5.214 & 0.780 & 1.345 & \multicolumn{2}{c}{-} \\
LGI, Finetuned LLSeg & 0.736 & 1.934 & 0.753 & 3.667 & 0.752 & 1.942 & \multicolumn{2}{c}{-} \\
\midrule
\multicolumn{9}{l}{\textbf{LinGuinE Configuration with Boosting}} \\
UGI, LLSeg & 0.753 & 1.865 & 0.662 & 6.088 & 0.632 & 5.111 & \textbf{0.491} & 3.620 \\
LGI, LLSeg & 0.755 & 1.738 & 0.672 & 5.511 & 0.773 & \textbf{1.198} & \multicolumn{2}{c}{-} \\
LGI, Finetuned LLSeg & \textbf{0.760} & \textbf{1.635} & 0.744 & 5.169 & 0.775 & 1.410 & \multicolumn{2}{c}{-}  \\
\bottomrule
\end{tabular}
\end{table}

LinGuinE configurations consistently outperform the pretrained LesionLocator model with respect to Dice on every dataset and ED on all lung datasets. While LesionLocator seemingly attains the best ED on APLCT, this metric is undefined when the model produces a false negative, which is a frequent cause of low Dice. This excludes the worst performing cases. Hence, we find LesionLocator to only be an appropriate choice when longitudinal training data is available. For comparison with Hering et al., we trained an nnUNet on the same datasets used for finetuning the guided models for lung cancer. We do not evaluate Hering et al. on APLCT due to lack of relevant data to train a suitable nnUNet. 

We found that for lung cancer datasets, using a segmentation algorithm finetuned for lung cancer with LGI in a LinGuinE context tends to yield the top results. This highlights the flexibility of LinGuinE that allows for performance enhancing, task-specific customisations. Nonetheless, as seen in Table \ref{results_table_median}, LinGuinE can outperform other methods by solely leveraging pretrained general-purpose algorithms like LLSeg, nnInteractive, and UGI - despite the nnUNet used for the Hering et al. comparison being designed specifically for lung cancer.

\subsection{Temporal Direction Agnosticism and Degradation}

For temporal analyses, we consider datasets with more than three timepoints per patient (Phase-3 and 4DCBCT) with their respective best configuration of LinGuinE based on Dice as in Table \ref{results_table_median}. To demonstrate that LinGuinE can propagate guidance in any temporal direction, we evaluated Dice distributions when using three different starting scans for each patient: (i) the earliest, (ii) the midpoint, and (ii) the latest. The set of segmentations for this three-way comparison is smaller than in other experiments, as we exclude any scan used as a starting point and any tumour not present in all starting scans for a patient. In Phase-3, 149 tumour segmentations from 77 scans across 61 patients were used. We found no statistically significant difference (three-way Friedmann chi-squared test, $p=0.4$) regardless of the source timepoint. Similar results were noted ($p=0.3$) in 4DCBCT, across 20 scans from 10 patients.

\begin{figure}[t]
    \centering
    \includegraphics[width=\columnwidth]{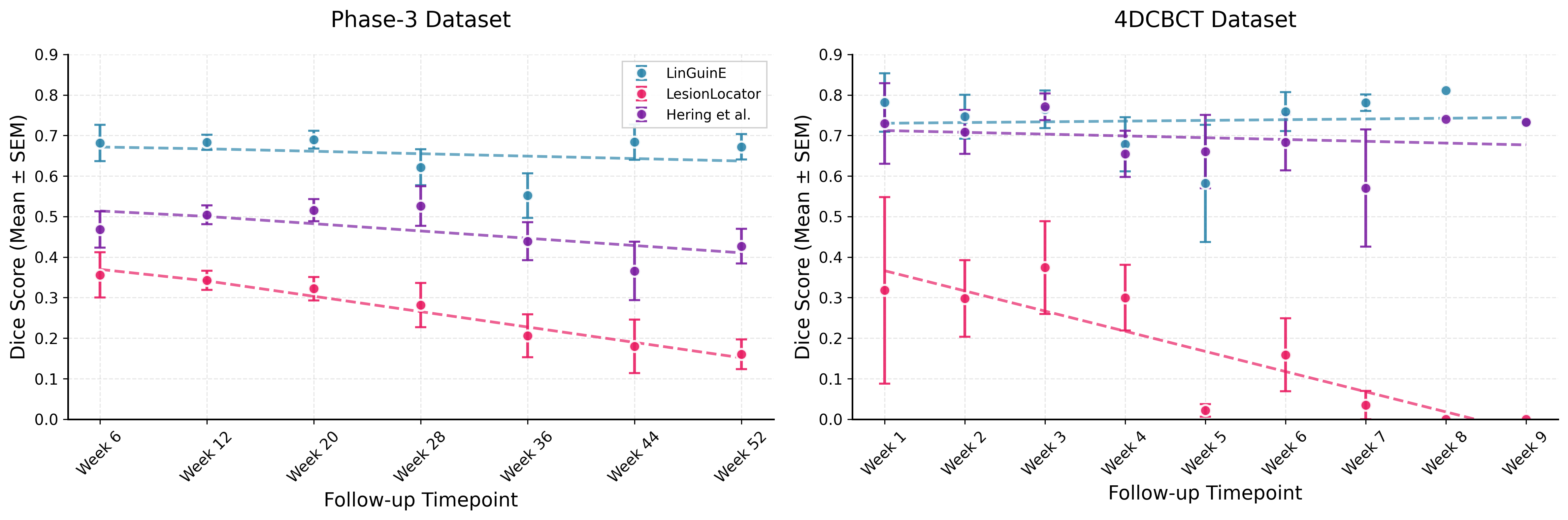}
    \caption{LinGuinE displays minimal performance degradation with elapsed time between the baseline and follow-up. Points represent mean Dice at various timepoints with the standard error on the mean displayed as error bars. The line of best fit for each method is shown for trend comparison - LinGuinE has the smallest slope.}
    \label{fig:time_degradation}
\end{figure}

To analyse temporal degradation, we consider Dice distributions for tumours at various timepoints. As seen in Figure \ref{fig:time_degradation}, there is minimal correlation between LinGuinE performance and the elapsed time between the baseline (used as the source) and the follow up, indicating robustness over the duration of the entire study compared to other methods, which display a clear downward trend. Across all patient studies, we find a mean decline in Dice of 2.591\% and 0.457\% per week in LesionLocator and Hering et al., respectively. LinGuinE displays a significantly smaller decline of 0.126\%. Moreover, the temporal direction agnosticism of our framework enables radiologists to mitigate this by selecting midpoint scans as the source, reducing the elapsed time between scan pairs as needed. 

\subsection{Ablation Studies}

\subsubsection{Radiologist Inputs} 
To assess the impact of propagating a point sampled from a mask versus a click, two radiologists semi-automatically re-segmented primary lung tumours at one random timepoint for 53 patients in the Phase-3 dataset. Radiologists were instructed to place at least one click on each tumour and were not informed of the experiment's aim, ensuring authentic interaction. We used those clicks as inputs into the best LinGuine configuration for Phase-3 to obtain 146 tumour segmentations in 132 scans on the remaining timepoints for those patients. The  Dice was 0.806. For the same tumours, when full ground truth masks were used to define $P_i$, it improved to 0.811. The improvement is statistically significant (paired t-test, $p=0.03$), suggesting that LinGuinE works best with a full segmentation as input. Nonetheless, the performance with a click input is similar to the reported inter-radiologist median Dice of 0.810 for manual lung tumour segmentation \cite{interobserver}.

\subsubsection{Impact of prompt boosting} As seen in Table \ref{results_table_median}, enabling prompt boosting yields LinGuinE configurations with the best ED metrics on three out of four datasets. The benefit is inconsistent—for example, in 4DCBCT the best performance was achieved without boosting. Conversely, in APLCT boosting improves Dice by 13.4\% and ED by 12.4\% in the best-performing configuration of UGI with LLSeg. We hypothesize that boosting is most beneficial when tumour tracking is challenging and registration struggles, but this requires further validation.
% While this requires further validation, we note clearer gains with the weakest tested registration algorithm (Arun's method); in Phase-3 with a finetuned LLSeg, it improves the Dice from 0.686 to 0.744. 

\subsubsection{Auto-regressive propagation} LesionLocator is the only prior method that analyzes an entire longitudinal study rather than a scan pair. Unlike LinGuinE, LesionLocator is autoregressive, propagating the previous timepoint prediction to account for changing tumour morphology. To evaluate the impact of auto-regression, we alter the LinGuinE workflow so that $P_{i}^{k}$ is first propagated to the $I_j$ closest in time to $I_i$. Then, $I_j$ becomes the source and $P_{j}^{k}$ is propagated to the next closest scan in time. $P_{j}^{k}$ is defined as the center of $M_j(k)$, where $M_j(k)$ is obtained via equation \eqref{guided_seg_destination}. Starting from the baseline scan and using the same algorithms as the best LinGuinE configuration, we compare this autoregressive workflow to the original in equation \eqref{full_linguine_eq} with paired t‑tests. In Phase-3, the autoregressive approach yields similar metrics (Dice=0.758, ED=1.620mm) with insignificant differences (Dice $p=0.3$, ED $p=0.1$). In 4DCBCT, we found a performance degradation (Dice=0.734, ED=4.447mm), which is statistically significant (Dice $p=0.02$, ED $p=0.003$). In UniToChest, we consider a subset of seven patients with more than two timepoints. The Dice drops from 0.746 to 0.741 (insignificant, $p=0.4$) and the ED remains at 0.983mm. 

Hence, we find no benefit in auto-regression. Our approach enables temporal direction agnosticism and in some scenarios avoids compounding errors by directly considering radiologist input before each propagation. We note that the temporal degradation observed for LesionLocator in Figure \ref{fig:time_degradation} differs from what was observed in the original paper \cite{LesionLocator}, and could possibly be attributable to the larger number of time points considered. By contrast, Hering et al. displays behaviour more akin to LinGuinE, and we hypothesize that this is because we invoked Hering et al. in a non-autoregressive manner. 

% appears to depend more on the number of follow-ups considered than elapsed time, suggesting that the decline may be attributable to auto-regression. 

\section{Conclusion}

In this study, we presented LinGuinE, a framework for longitudinal semi-automatic volumetric segmentation and tracking of tumours without the requirement of training on longitudinal data. LinGuinE demonstrates accurate segmentation on four unseen datasets with a total of 456 longitudinal studies, outperforming other methods. Our method scales to an arbitrary number of timepoints, is temporally agnostic to the initial scan, and segments the tumour boundaries with less temporal degradation than existing approaches. LinGuinE can potentially be applied as a facilitating tool for the introduction of volumetric biomarkers as a response criterion in oncological therapy. We release LinGuinE as a PyTorch package to support this objective and to assist in the creation of annotated longitudinal datasets with tracked tumours. Further work is needed to handle disappearing or coalescing lesions, and to establish LinGuinE as a segmentation tool for different cancer types and imaging modalities.

 %% removed for anonymized MICCAI submission.
    
    % The following acknowledgement and disclaimer sections can be removed for the double-blind review process.  If and when your paper is accepted, reinsert the acknowledgement and the disclaimer clause in your final camera-ready version.
    % IF you opted to include the acknowledgement and disclaimer sections, they will count towards the 8-page limit.

% \begin{credits}
% \subsubsection{\ackname} A bold run-in heading in small font size at the end of the paper is
% used for general acknowledgments, for example: This study was funded

% \subsubsection{\discintname}
% It is now necessary to declare any competing interests or to specifically
% state that the authors have no competing interests. Please place the
% statement with a bold run-in heading in small font size beneath the
% (optional) acknowledgments\footnote{If EquinOCS, our proceedings submission
% system, is used, then the disclaimer can be provided directly in the system.},
% for example: The authors have no competing interests to declare that are
% relevant to the content of this article. Or: Author A has received research
% grants from Company W. Author B has received a speaker honorarium from
% Company X and owns stock in Company Y. Author C is a member of committee Z.
% \end{credits}


\bibliographystyle{splncs04}
\bibliography{core_ref}

\end{document}